\documentclass[preprint,showpacs,preprintnumbers,amsmath,amssymb]{revtex4}

\usepackage{graphicx}
\usepackage{dcolumn}
\usepackage{bm}
\usepackage{hyperref}
\usepackage{array}
\hypersetup{colorlinks=False}

\begin{document}
\newcommand{\RR}[1]{[#1]}
\newcommand{\intsum}{\sum \kern -15pt \int}
\newfont{\Yfont}{cmti10 scaled 2074}
\newcommand{\Y}{\hbox{{\Yfont y}\phantom.}}
\def\O{{\cal O}}
\newcommand{\bra}[1]{\left< #1 \right| }
\newcommand{\braa}[1]{\left. \left< #1 \right| \right| }
\def\Bra#1#2{{\mbox{\vphantom{$\left< #2 \right|$}}}_{#1}
\kern -2.5pt \left< #2 \right| }
\def\Braa#1#2{{\mbox{\vphantom{$\left< #2 \right|$}}}_{#1}
\kern -2.5pt \left. \left< #2 \right| \right| }
\newcommand{\ket}[1]{\left| #1 \right> }
\newcommand{\kett}[1]{\left| \left| #1 \right> \right.}
\newcommand{\scal}[2]{\left< #1 \left| \mbox{\vphantom{$\left< #1 #2 \right|$}}
\right. #2 \right> }
\def\Scal#1#2#3{{\mbox{\vphantom{$\left<#2#3\right|$}}}_{#1}
{\left< #2 \left| \mbox{\vphantom{$\left<#2#3\right|$}} \right. #3
\right> }}


\title{Photon asymmetries in $nd\rightarrow$ $^3H\gamma$ using EFT($\pi\!\!\!/$) approach}

\author{M. Moeini Arani}%
\email[]{m.moeini.a@khayam.ut.ac.ir (corresponding author)}
\affiliation {Department of Physics, University of Tehran, P.O.Box
14395-547, Tehran, Iran }

\author{S. Bayegan }
\email[]{bayegan@khayam.ut.ac.ir} \affiliation {Department of
Physics, University of Tehran, P.O.Box 14395-547, Tehran, Iran }%

\date{\today}

\begin{abstract}
The parity-violating Lagrangian of the weak nucleon-nucleon ($NN$)
interaction in the pionless effective field theory
(EFT($/\!\!\!\pi$)) approach contains five independent unknown
low-energy coupling constants (LECs). The photon asymmetry with
respect to neutron polarization in $np\rightarrow$ $d\gamma$
$A^{np}_\gamma$, the circular polarization of outgoing photon in
$np\rightarrow$ $d\gamma$ $P^{np}_\gamma$, the neutron spin rotation
in hydrogen $\frac{1}{\rho}\frac{d\phi^{np}}{dl}$, the neutron spin
rotation in deuterium $\frac{1}{\rho}\frac{d\phi^{nd}}{dl}$ and the
circular polarization of $\gamma$-emission in $nd\rightarrow$
$^3H\gamma$ $P^{nd}_\gamma$ are the parity-violating observables
which have been recently calculated in terms of parity-violating
LECs in the EFT($/\!\!\!\pi$) framework. We obtain the LECs by
matching the parity-violating observables to the Desplanques,
Donoghue, and Holstein (DDH) best value estimates. Then, we evaluate
photon asymmetry with respect to the neutron polarization
$a^{nd}_\gamma$ and the photon asymmetry in relation to deuteron
polarization $A^{nd}_\gamma$ in $nd\rightarrow$ $^3H\gamma$ process.
We finally compare our EFT($/\!\!\!\pi$) photon asymmetries results
with the experimental values and the previous calculations based on
the DDH model.
\end{abstract}

\maketitle

\section{Introduction}\label{sec:introduction}
The theoretical analyses of the hadronic parity violation in
few-nucleon systems were introduced by Danilov \cite{Danilov} and
Desplanques and Missimer \cite{Despl-miss} for the first time . The
analysis of parity-violating (PV) couplings in a meson-exchange
model was then developed by Desplanques, Donoghue, and Holstein
(DDH) \cite{DDH}. At low energies, the DDH potential is dominated by
the exchange of light mesons as a sign of the long-distance physics.
The inclusion of heavy $\rho$ and $\omega$ mesons in DDH model point
out the short-distance physics. With these model assumptions, the
DDH potential has been widely used to calculate PV observables in
terms of PV meson-exchange couplings for different few-body systems
\cite{Review of PV}.

The idea of the effective field theory (EFT) and its application in
low-energy parity violation is principally based on the fact that
the long-distance physics should be independent of the details of
the short-distance effects. Therefore, the PV transition amplitudes
with no reference to the description of the short-distance effects
are appropriate in order to determine the PV observables.

Recently, the PV Lagrangian of the weak nucleon-nucleon ($NN$)
interaction of the pionless effective field theory (EFT($\pi\!\!\!\!
/$)) with five independent unknown low-energy coupling constants
(LECs) has been incorporated to evaluate the following PV
observables in terms of PV LECs in a unified EFT($\pi\!\!\!\! /$)
framework. The photon asymmetry with respect to neutron polarization
$A^{np}_\gamma$ and the circular polarization of outgoing photon
$P^{np}_\gamma$ in $np\rightarrow$ $d\gamma$ were calculated by
Schindler $\textit{et al.}$ \cite{schindler-springer}. The neutron
spin rotation in hydrogen $\frac{1}{\rho}\frac{d\phi^{np}}{dl}$ and
the neutron spin rotation in deuterium
$\frac{1}{\rho}\frac{d\phi^{nd}}{dl}$ were evaluated by Griesshammer
$\textit{et al.}$ in \cite{G-S-S}. Also, the circular polarization
of $\gamma$-emission in $nd\rightarrow$ $^3H\gamma$ $P^{nd}_\gamma$
was calculated by Arani and Bayegan in \cite{moeini-bayegan}.

The older experimental data for $A^{np}_\gamma$ are $(-0.6\pm0.21
)\times10^{-7}$ \cite{C-V-W} and $(-0.15\pm 0.47)\times10^{-7}$
\cite{22 of snow}. The newer measurement of $A^{np}_\gamma$ has been
reported with a statistical precision of $5\times10^{-9}$ and with
negligible systematic error \cite{snow et al}. The only available
data for $P^{np}_\gamma$ is $(1.8\pm1.8)\times10^{-7}$
\cite{Knyazkov et al}. At present, there are no experimental data on
the neutron-proton ($np$) and the neutron-deuteron ($nd$) spin
rotations. The experimental data for $P^{nd}_\gamma$ has not been
reported to this point. In order to determine PV LECs, we need to
increase the accuracy of existing measurement and to plan new
experiments for measuring the neutron spin rotation on a variety of
targets and the circular photon polarization for unpolarized beam
and target \cite{Review of PV}.

The lack of experimental values for the PV observables are the main
obstacle to determine the PV LECs. Therefore, in this paper, we
intend to obtain the five independent PV LECs by matching the five
EFT($\pi\!\!\!\! /$) relations for PV observabes $A^{np}_\gamma$,
$P^{np}_\gamma$, $\frac{1}{\rho}\frac{d\phi^{np}}{dl}$,
$\frac{1}{\rho}\frac{d\phi^{nd}}{dl}$ and $P^{nd}_\gamma$ to DDH
"best" values estimates for these observables. The determination of
the PV LECs provides the calculation of the asymmetry of the
outgoing photon with respect to the neutron $a^{nd}_\gamma$, and the
deuteron $A^{nd}_\gamma$, polarizations. The photon asymmetries
results are then compared with the experimental values and the
previous calculations based on the DDH model.

The reminder of this paper is organized as follows. In
Sect.\ref{sec:Lagrangian} we briefly review the leading-order (LO)
Lagrangians of the strong two- and three-nucleon, the weak $NN$ and
the electromagnetic interactions. In Sect.\ref{nd radiative capture
syatem}, we introduce the parity-conserving (PC) and
parity-violating amplitudes of $nd\rightarrow$ $^3H\gamma$ and
calculate the $P^{nd}_\gamma$, $a^{nd}_\gamma$ and
 $A^{nd}_\gamma$ observables in $nd\rightarrow$
$^3H\gamma$ process in terms of PV LECs. In Sect.\ref{PV LECs} the
EFT($\pi\!\!\!\! /$) results of $A^{np}_\gamma$, $P^{np}_\gamma$,
$\frac{1}{\rho}\frac{d\phi^{np}}{dl}$ and
$\frac{1}{\rho}\frac{d\phi^{nd}}{dl}$ observables in terms of the PV
low-energy constants are presented and used for obtaining PV LECs.
Sect.\ref{Ag calculation} contains our EFT($\pi\!\!\!\! /$) results
for $a^{nd}_\gamma$ and
 $A^{nd}_\gamma$ which are evaluated by the determined PV LECs. The conclusion and outlook are expressed in
Sect.\ref{conclusion}.

\section{Lagrangian}\label{sec:Lagrangian}
In this section, we introduce Lagrangians of the strong, weak and
electromagnetic interactions in the three-body system at the lowest
order. In the EFT approach with the Z-parametrization structure, the
strong two- and three-nucleon interactions are presented by
\cite{phillips-rupak-savage,20 of sadeghi-bayegan}
\begin{eqnarray}\label{Eq:1}
\mathcal{L}^{PC}=N^\dag\Big(iD_0+\frac{\vec{D}^2}{2m_N}\Big)N+d^{A^\dag}_{s}\Big[\Delta_s-c_{0s}\Big(iD_0+\frac{\vec{D}^2}{4m_N}+\frac{\gamma^2_s}{m_N}\Big)\Big]d^{A}_{s}\qquad\;\;
  \qquad\qquad\qquad\qquad\;\;\;\nonumber \\
  +d^{i^\dag}_{t}\Big[\Delta_t-c_{0t}\Big(iD_0+\frac{\vec{D}^2}{4m_N}+\frac{\gamma^2_t}{m_N}\Big)\Big]d^{i}_{t}
  -y\Big(d^{A^\dag}_{s}(N^T P^A N)+d^{i^\dag}_{t}(N^T P^i
  N)+h.c.\Big)\qquad\quad\!\nonumber\\ +\frac{m_N y^2
  H_0(\Lambda)}{3\Lambda^2}\;N^\dag\bigg(\big(d^{i}_{t}\sigma_i\big)^\dag\big(d^{j}_{t}\sigma_j\big)-\big[\big(d^{i}_{t}\sigma_i\big)^\dag\big(d^{A}_{s}\tau_A\big)
  +h.c.\big]+\big(d^{A}_{s}\tau_A\big)^\dag\big(d^{B}_{s}\tau_B\big)\bigg)N+...\,,
\end{eqnarray}
where $N$ is the nucleon iso-doublet and the auxiliary fields
$d^{A(B)}_{s}$ and $d^{i(j)}_{t}$ carry the quantum numbers of
$^1S_0$ di-nucleon and the deuteron, respectively. $D_\mu =
\partial_\mu + ie\frac{1+\tau_3}{2}A_\mu$ is the nucleon covariant derivative. The
operators $P^{i}$ and $P^{A}$ project $NN$ system into the triplet
and singlet channels, respectively. $\tau_{A/B}$ ($\sigma_{i/j}$)
are isospin (spin) pauli matrices with $A,B=1,2,3$ ($i,j=1,2,3$) as
iso-triplet (vector) indices. $m_N$ is the nucleon mass and the
three-nucleon interaction at the leading order is $H_0(\Lambda)$
with cut-off $\Lambda$. The parameters of the Lagrangian in
Eq.(\ref{Eq:1}) are fixed using Z-parametrization \cite{20 of
sadeghi-bayegan}. So, dibaryon-nucleon-nucleon ($dNN$) coupling
constant is chosen as $y^2=\frac{4\pi}{m_N}$ and the LO parameters
$\Delta_{s/t}$ are obtained from the poles of the $NN$ S-wave
scattering amplitude at $i\gamma_{s/t}$ where
$\gamma_s=\frac{1}{a_s}$ with $a_s$ as the scattering length in
$^1S_0$ state and $\gamma_t$ is the binding momentum of the
deuteron.

The Lagrangian
\begin{eqnarray}\label{Eq:04}
{\mathcal{L}_{PV}}=-\Big[g^{(^3S_1-^1P_1)}d^{i^\dag}_{t}N^T
   i\left(\overleftarrow{\nabla}\sigma_2\tau_2-\sigma_2\tau_2\overrightarrow{\nabla}\right)_iN\qquad\qquad\qquad\quad\;\;
 \nonumber \\
+g^{(^1S_0-^3P_0)}_{(\Delta I=0)}d^{A^\dag}_{s}N^T
i\left(\overleftarrow{\nabla}\sigma_2\sigma_i\tau_2\tau_A-\sigma_2\sigma_i\tau_2\tau_A\overrightarrow{\nabla}\right)_iN\qquad\;\;
\nonumber \\
+g^{(^1S_0-^3P_0)}_{(\Delta I=1)}\epsilon^{3AB}d^{A^\dag}_{s}\,N^T
i\left(\overleftarrow{\nabla}\sigma_2\sigma_i\tau_2\tau^B-\sigma_2\sigma_i\tau_2\tau^B\overrightarrow{\nabla}\right)_iN
\nonumber \\
+g^{(^1S_0-^3P_0)}_{(\Delta I=2)}\mathcal{I}^{AB}d^{A^\dag}_{s}N^T
i\left(\overleftarrow{\nabla}\sigma_2\sigma_i\tau_2\tau^B-\sigma_2\sigma_i\tau_2\tau^B\overrightarrow{\nabla}\right)_iN\,\,
\nonumber \\
+g^{(^3S_1-^3P_1)}\epsilon^{ijk}d^{i^\dag}_{t}N^T
\left(\overleftarrow{\nabla}\sigma_2\sigma^k\tau_2\tau_3-\sigma_2\sigma^k\tau_2\tau_3\overrightarrow{\nabla}\right)^jN\Big]\quad
\nonumber \\
+h.c.+...
\,.\qquad\qquad\qquad\qquad\qquad\qquad\qquad\qquad\qquad\qquad
\end{eqnarray}
introduces the weak $\textit{NN}$ interactions with the dibaryon
formalism at the very-low-energy regime. The Lagrangian in
Eq.(\ref{Eq:04}) is constructed by considering five transitions
which mix the S-P partial waves. In the above equation, the weak
$dNN$ coupling constant for the PV $\textit{NN}$ transition between
$\bar{X}$ and $\bar{Y}$ partial waves is depicted by
$g^{(\bar{X}-\bar{Y})}$. $\Delta I$ represents the isospin change in
the PV vertex and

\begin{eqnarray}\label{Eq:5}
  \mathcal{I}=\left(
                \begin{array}{ccc}
                  1 & 0 & 0 \\
                  0 & 1 & 0 \\
                  0 & 0 & -2 \\
                \end{array}
              \right).
\end{eqnarray}

With considering the PV $\textit{NN}$ Lagrangian introduced in
Eq.(\ref{Eq:04}), the parity-violating observables in the three-body
systems are cutoff independent at the leading order and
next-to-leading order (NLO)
\cite{Grieshammer-schindler,moeini-bayegan}. So, the
parity-violating three-nucleon interaction (PV 3NI) is not
participated in the three-nucleon systems at LO and NLO in
EFT($/\!\!\!\pi$).

At the leading order, the electromagnetic interactions are
introduced by the Lagrangian
\begin{eqnarray}\label{Eq:6}
  \mathcal{L}^{EM}=\frac{e}{2m_N}\bigg[N^\dag\big(k_0+k_1\tau^3\big)\vec{\sigma}\cdot\vec{B}N+N^\dag\frac{(1+\tau_3)}{2}\:(\vec{P}+\vec{P^\prime})\cdot\vec{\varepsilon_\gamma^\ast}N\bigg],
\end{eqnarray}
where the first and second terms indicate the M1 and E1 interactions
of a photon with a single nucleon, respectively. $k_0$ and $k_1$ are
the isoscalar and isovector nucleon magnetic moments. $e$,
$\vec{B}$, $\vec{P}$ $(\vec{P^\prime})$ and
$\vec{\varepsilon}_\gamma$ denote the electric charge, the magnetic
field, the incoming (outgoing) nucleon momentum and the 3-vector
polarization of the produced photon, respectively.

\section{$nd\rightarrow$ $^3H\gamma$ system}\label{nd radiative capture syatem}

\subsection{PC amplitude of $nd\rightarrow$ $^3H\gamma$ process}\label{PC nd radiative capture}

\begin{figure*}[tb]\centering
\includegraphics*[width=12cm]{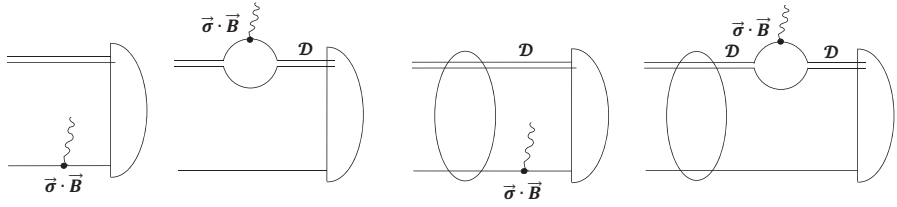}
\caption{\label{Fig:PC nd capture}The leading-order diagrams which
contribute to PC (M1) amplitude of $nd\rightarrow$ $^3H\gamma$.
Single, double and wavy lines denote a nucleon, a dibaryon auxiliary
field with the propagator matrix of $\mathcal{D}=diag(D_{t},D_{s})$
and a photon, respectively. Dashed oval and dashed half-oval
indicate $Nd$ scattering amplitude at LO which is shown in
Fig.\ref{Fig:PC nd scattering} of Appendix and the formation of
triton (the normalized triton wave function).}
\end{figure*}

The leading-order diagrams which contribute to the PC amplitude of
$nd\rightarrow$ $^3H\gamma$ process are shown in Fig.\ref{Fig:PC nd
capture}. In the cluster-configuration space the propagators of two
intermediate dibaryon auxiliary fields $d^i_{t}$ and $d^A_{s}$ at LO
are presented by $\mathcal{D}=diag(D_{t},D_{s})$ where $D_{t(s)}$ is
\begin{eqnarray}\label{Eq:7}
  D_{t(s)}(q_0,q)=\frac{1}{\gamma_{t(s)}-\sqrt{\frac{q^2}{4}-m_Nq_0-i\varepsilon}}\,.
\end{eqnarray}
The PC nucleon-deuteron ($Nd$) scattering (the dashed oval) and the
normalized triton wave function (the dashed half-oval) in
Fig.\ref{Fig:PC nd capture} are obtained by solving the Faddeev
equation and the homogenous part of the Faddeev equation introduced
in Fig.\ref{Fig:PC nd scattering} of Appendix, respectively \cite{20
of sadeghi-bayegan,moeini-bayegan}.

 In order to compute the LO PC amplitude of $nd\rightarrow$ $^3H\gamma$ process
for two M1 transitions corresponding to the initial
$^2S_{\frac{1}{2}}$ and $^4S_{\frac{3}{2}}$ states, we use the
Lagrangian of Eq.(\ref{Eq:1}) for the strong interaction and the
first terms of Eq.(\ref{Eq:6}) for the M1 interaction. In the
cluster-configuration space, one can be able to write the
contribution of the diagrams in Fig.\ref{Fig:PC nd capture} for both
possible magnetic transitions with initial doublet and quartet
channels as
\begin{eqnarray}\label{Eq:008}
  \mathcal{M}^{PC}=\mathcal{W}^{PC}(^2S_{\frac{1}{2}})Y^{PC}(^2S_{\frac{1}{2}})+\mathcal{W}^{PC}(^4S_{\frac{3}{2}})Y^{PC}(^4S_{\frac{3}{2}}),\:
\end{eqnarray}
where
\begin{eqnarray}\label{Eq:8}
  Y^{PC}(^2S_{\frac{1}{2}})=t^\dag\big(i\vec{\varepsilon}_d\cdot\vec{\varepsilon}^\ast_\gamma\times\vec{\tilde{q}}+\vec{\sigma}\times\vec{\varepsilon}_d\cdot\vec{\varepsilon}^\ast_\gamma\times\vec{\tilde{q}}\big)N\,,\!
\nonumber\\
Y^{PC}(^4S_{\frac{3}{2}})=t^\dag\big(2i\vec{\varepsilon}_d\cdot\vec{\varepsilon}^\ast_\gamma\times\vec{\tilde{q}}-\vec{\sigma}\times\vec{\varepsilon}_d\cdot\vec{\varepsilon}^\ast_\gamma\times\vec{\tilde{q}}\big)N\,.
\end{eqnarray}
$t$, $\vec{\varepsilon}_d$ and $\vec{\tilde{q}}$ are the final $^3H$
(or $^3He$) field, the 3-vector polarization of the deuteron and the
unit vector along the 3-momentum of the photons, respectively. In
Eq.(\ref{Eq:008}), $\mathcal{W}^{PC}(X)$ with
$X=^2$$S_{\frac{1}{2}},\,^4S_{\frac{3}{2}}$ is a $2\times1$ matrix
which is given in the cluster-configuration space by
\begin{eqnarray}\label{Eq:08}
\mathcal{W}^{PC}(X)=\left(
                             \begin{array}{c}
                               \mathcal{W}^{PC}_{nd_t\rightarrow ^3H_{(nd_t)}\gamma}(X) \\
                               \mathcal{W}^{PC}_{nd_t\rightarrow ^3H_{(nd_s)}\gamma}(X) \\
                             \end{array}
                           \right)
\end{eqnarray}
where $\mathcal{W}^{PC}_{nd_t\rightarrow^3H_{(nd_{t(s)})}\gamma}(X)$
denotes the contribution of all diagrams in Fig.\ref{Fig:PC nd
capture} for PC $nd_t\rightarrow ^3$$H_{(nd_{t(s)})}\gamma$
transition. $^3$$H_{(nd_{t})}$ and $^3$$H_{(nd_{s})}$ represent the
final triton components which are made from the final doublet $nd_t$
(nucleon and triplet dibaryon) and $nd_s$ (nucleon and singlet
dibaryon) cases, respectively.

The results of $\mathcal{W}^{PC}(X)$
($X=\,$$^2S_{\frac{1}{2}},\:^4S_{\frac{3}{2}}$) are used for the
evaluation of the PV observables in Sect.\ref{PV observables in nd
radiative capture}.

\subsection{PV amplitude of $nd\rightarrow$ $^3H\gamma$ process}\label{PV nd radiative capture}

\begin{figure*}[tb]\centering
\includegraphics*[width=15cm]{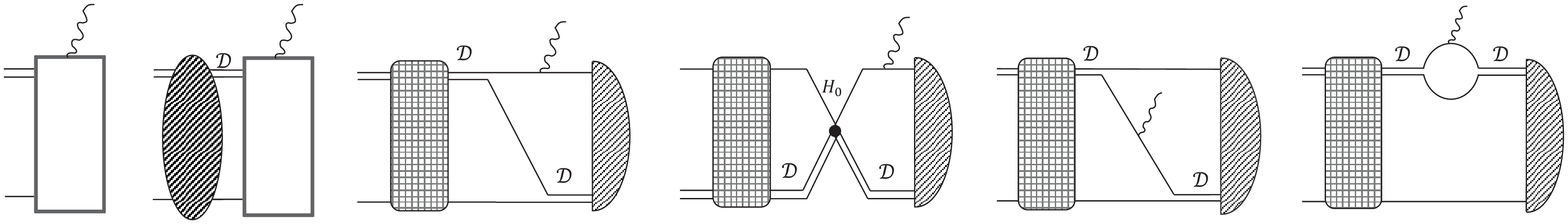} \caption{\label{Fig:PV Nd capture} The diagrams of the PV $nd\rightarrow$ $^3H\gamma$ process at LO.
The photon-nucleon-nucleon vertex represents one-body current for E1
interaction. $H_0$ is three-body interaction which renormalizes the
$Nd$ scattering amplitude at LO. The dashed rectangular and the box
with the wavy line are the PV $Nd$ scattering amplitude and the sum
of several diagrams which participate in the PV $nd\rightarrow$
$^3H\gamma$ amplitude at LO, respectively (for more detail see
Appendix). All notations are the same as the previous figures.}
\end{figure*}

In the next step, we explain briefly our EFT($\pi\!\!\!/$) results
for the PV (E1) amplitude of $nd\rightarrow$ $^3H\gamma$ process
based on the procedure presented in ref.\cite{moeini-bayegan}.

The Lagrangians in Eqs.(\ref{Eq:04}) and (\ref{Eq:6}) are used for
the weak and E1 interactions, respectively. By working in the
coulomb gauge, the spin structure of the PV (E1) amplitude of
$nd\rightarrow$ $^3H\gamma$ process can be written as two orthogonal
terms,
\begin{eqnarray}\label{Eq:9}
  i\big(t^\dag\sigma_aN\big)\big(\vec{\varepsilon}_d\times\vec{\varepsilon}^\ast_\gamma\big)_a\:,\qquad \big(t^\dag
  N\big)\big(\vec{\varepsilon}_d\cdot\vec{\varepsilon}^\ast_\gamma\big)\,.\;\;\;
\end{eqnarray}
Similar to the PC subsection, the contribution of E1 transition
which is the sum of both contributions corresponding to the incoming
doublet ($X=^2$$S_{\frac{1}{2}}$) and quartet
($X=^4$$S_{\frac{3}{2}}$) channels can be written as
\begin{eqnarray}\label{Eq:010}
  \mathcal{M}^{PV}=\mathcal{W}^{PV}(^2S_{\frac{1}{2}})Y^{PV}(^2S_{\frac{1}{2}})+\mathcal{W}^{PV}(^4S_{\frac{3}{2}})Y^{PV}(^4S_{\frac{3}{2}}),\:
\end{eqnarray}
with
\begin{eqnarray}\label{Eq:10}
  Y^{PV}(^2S_{\frac{1}{2}})=\;t^\dag\big(\vec{\varepsilon}_d\cdot\vec{\varepsilon}^\ast_\gamma+i\vec{\sigma}\cdot\vec{\varepsilon}_d\times\vec{\varepsilon}^\ast_\gamma\big)N\,,
\nonumber\\
Y^{PV}(^4S_{\frac{3}{2}})=\;t^\dag\big(\vec{\varepsilon}_d\cdot\vec{\varepsilon}^\ast_\gamma+i\vec{\sigma}\cdot\vec{\varepsilon}_d\times\vec{\varepsilon}^\ast_\gamma\big)N\,,
\end{eqnarray}
where $\mathcal{W}^{PV}(X)=\left(
                             \begin{array}{c}
                               \mathcal{W}^{PV}_{nd_t\rightarrow ^3H_{(nd_t)}\gamma}(X) \\
                               \mathcal{W}^{PV}_{nd_t\rightarrow ^3H_{(nd_s)}\gamma}(X) \\
                             \end{array}
                           \right)$ with $\mathcal{W}^{PV}_{nd_t\rightarrow
^3H_{(nd_{t(s)})}\gamma}(X)$ denotes the contribution of PV
$nd_t\rightarrow ^3H_{(nd_{t(s)})}\gamma$ amplitude for E1
transition in the cluster-configuration space with the incoming $X$
partial wave.

All possible diagrams which consider the weak interaction effects in
$nd\rightarrow$ $^3H\gamma$ process at LO are depicted in
Fig.\ref{Fig:PV Nd capture}. The dashed rectangular with solid line
around it denotes the LO PV $Nd$ scattering amplitude which is
recently calculated by Griesshammer $\textit{et al.}$, with EFT
approach \cite{G-S-S}. The box with the wavy line is the sum of
several diagrams which contribute in the amplitude of PV radiative
capture of neutron by deuteron. All diagrams which make
contributions of the dashed rectangular (PV $Nd$ scattering
amplitude) and the box with the wavy line in Fig.\ref{Fig:PV Nd
capture} are presented briefly in Appendix.

With respect to the diagrams introduced in Fig.\ref{Fig:PV Nd
capture}, we obtain the PV amplitude of neutron radiative capture by
deuteron for incoming $X$ ($X=$$^2S_{\frac{1}{2}}$,
$^4S_{\frac{3}{2}}$) channel in terms of the PV LECs which have been
introduced in Eq.(\ref{Eq:04}),
\begin{eqnarray}\label{Eq:11}
\mathcal{W}^{PV}(X)=\;\;a_{_X}(^3S_1-^1P_1)\,g^{(^3S_1-^1P_1)}\qquad\quad\;\;
\nonumber\\+a_{_X}(^1S_0-^3P_0,\Delta
I=0)\,g^{(^1S_0-^3P_0)}_{(\Delta I=0)}
\nonumber\\
+a_{_X}(^1S_0-^3P_0,\Delta I=1)\,g^{(^1S_0-^3P_0)}_{(\Delta I=1)}
\nonumber\\+a_{_X}(^3S_1-^3P_1)\,g^{(^3S_1-^3P_1)}.\qquad\quad\,\,
\end{eqnarray}
In Eq.(\ref{Eq:11}), the $a_{_X}(\bar{X}-\bar{Y})$ is a $2\times 1$
coefficient matrix of $g^{(\bar{X}-\bar{Y})}$ for incoming $X$
partial wave. We evaluate numerically the value of the
$a_{_X}(\bar{X}-\bar{Y})$ coefficient matrix for
$X=^2$$S_{\frac{1}{2}},^4$$S_{\frac{3}{2}}$ with
$\Lambda=700\:\textrm{MeV}$. Our results for
$a_{_X}(\bar{X}-\bar{Y})$ coefficient matrix are shown in Table
\ref{tab:ax}. The leading-order $\mathcal{W}^{PV}(X)$ does not
contain $g^{(^1S_0-^3P_0)}_{(\Delta I=2)}$ term because the $Nd$
system is an iso-doublet and PV coupling $g^{(^1S_0-^3P_0)}_{(\Delta
I=2)}$ cannot contribute.
\begin{table*}[tb]\centering
\caption{Results of $a_{_X}(\bar{X}-\bar{Y})$ for the different
incoming and outgoing partial waves with
$\Lambda=700\;\textrm{MeV}$.
$a_{_{^2S_{\frac{1}{2}}}}(\bar{X}-\bar{Y})$ and
$a_{_{^4S_{\frac{3}{2}}}}(\bar{X}-\bar{Y})$ results are in
$10^{-2}\,\textrm{MeV}^{-1}$ and $10^{-4}\,\textrm{MeV}^{-1}$ units,
respectively.}
\label{tab:ax}       
\begin{tabular}{lllll}\hline\noalign{\smallskip}
 $X$ & $a_{_X}(^3S_1-^1P_1)$ &
$a_{_X}(^1S_0-^3P_0,\Delta I=0)$ &$a_{_X}(^1S_0-^3P_0,\Delta I=1)$
&$a_{_X}(^3S_1-^3P_1)$  \\
\hline\hline  \noalign{\smallskip}

$^2S_{\frac{1}{2}}$ & $\left(
                        \begin{array}{c}
                          -0.75 \\
                          0.55 \\
                        \end{array}
                      \right)$
 & $\left(
                        \begin{array}{c}
                          1.24 \\
                          -0.80 \\
                        \end{array}
                      \right)$& $\left(
                        \begin{array}{c}
                          -0.73 \\
                          0.49 \\
                        \end{array}
                      \right)$& $\left(
                        \begin{array}{c}
                          2.35 \\
                          -1.16 \\
                        \end{array}
                      \right)$ \\

$^4S_{\frac{3}{2}}$ & $\left(
                        \begin{array}{c}
                          7.62 \\
                          -3.10 \\
                        \end{array}
                      \right)$ & $\left(
                        \begin{array}{c}
                          3.32 \\
                          -0.63 \\
                        \end{array}
                      \right)$& $\left(
                        \begin{array}{c}
                          0.42 \\
                          -0.26 \\
                        \end{array}
                      \right)$ & $\left(
                        \begin{array}{c}
                          -28.87 \\
                          7.93 \\
                        \end{array}
                      \right)$ \\


\noalign{\smallskip}\hline
\end{tabular}
\end{table*}
\begin{table*}[tb]\centering
\caption{Results of $Abs[1-\frac{a_{_X}(\bar{X}-\bar{Y})\; at\;
\Lambda=400\;\textrm{MeV}}{a_{_X}(\bar{X}-\bar{Y})\; at\;
\Lambda=700\;\textrm{MeV}}]$ for the different incoming and outgoing
partial waves.}
\label{tab:c.v. of ax}       
\begin{tabular}{lllll}
\hline\noalign{\smallskip}

X & $(^3S_1-^1P_1)$ & $(^1S_0-^3P_0,\Delta I=0)$  &
$(^1S_0-^3P_0,\Delta I=1)$
& $(^3S_1-^3P_1)$  \\
\hline\hline \noalign{\smallskip}

$^2S_{\frac{1}{2}}$ & $\left(
                        \begin{array}{c}
                          0.0477 \\
                          0.0978 \\
                        \end{array}
                      \right)$ & $\left(
                        \begin{array}{c}
                          0.0358 \\
                          0.0917 \\
                        \end{array}
                      \right)$& $\left(
                        \begin{array}{c}
                          0.0526 \\
                          0.1098 \\
                        \end{array}
                      \right)$ & $\left(
                        \begin{array}{c}
                          0.0212 \\
                          0.0591 \\
                        \end{array}
                      \right)$ \\
$^4S_{\frac{3}{2}}$ & $\left(
                        \begin{array}{c}
                          0.0300 \\
                          0.1100 \\
                        \end{array}
                      \right)$ & $\left(
                        \begin{array}{c}
                          0.1402 \\
                          0.1474 \\
                        \end{array}
                      \right)$& $\left(
                        \begin{array}{c}
                          0.1057 \\
                          0.0888 \\
                        \end{array}
                      \right)$ & $\left(
                        \begin{array}{c}
                          0.0021 \\
                          0.0350 \\
                        \end{array}
                      \right)$ \\

\noalign{\smallskip}\hline\\
\end{tabular}
\end{table*}

The cutoff variation of the $a_{_X}(\bar{X}-\bar{Y})$ results are
investigated by calculating the
$Abs[1-\frac{a_{_X}(\bar{X}-\bar{Y})\; at\;
\Lambda=400\;\textrm{MeV}}{a_{_X}(\bar{X}-\bar{Y})\; at\;
\Lambda=700\;\textrm{MeV}}]$ for the different incoming and outgoing
partial waves. Our results for the cutoff variation of the
coefficient matrix $a_{_X}(\bar{X}-\bar{Y})$ are shown in Table
\ref{tab:c.v. of ax}. These results indicate that the cutoff
dependence is small at the leading order. This small variation can
be removed with the consideration of the higher-order corrections.
The negligible cutoff variation of the PV amplitude of
$nd\rightarrow$ $^3H\gamma$ process represents that we do not need
the PV 3NI at LO because the PV amplitude of the neutron radiative
capture by deuteron is properly renormalized.

In the next sections, we concentrate on finding the PV observables
in $nd\rightarrow$ $^3H\gamma$ process with using Eq.(\ref{Eq:11})
and the results in the Table \ref{tab:ax}.

\subsection{PV observables in $nd\rightarrow$ $^3H\gamma$ process in terms of PV LECs}\label{PV observables in nd radiative capture}
We have briefly clarified the PC and PV amplitudes of
$nd\rightarrow$ $^3H\gamma$ process, we then proceed by introducing
the PV observables in $nd\rightarrow$ $^3H\gamma$. The circular
polarization of outgoing photon $P^{nd}_\gamma$, the photon
asymmetry with respect to the neutron polarization $a^{nd}_\gamma$
and the asymmetry of photon in relation to the deuteron polarization
$A^{nd}_\gamma$ are three PV observables in $nd\rightarrow$
$^3H\gamma$. We discuss about the calculation of these observables
in the following.

\subsubsection{Photon circular polarization in $nd\rightarrow$ $^3H\gamma$}\label{Pg}
The PV polarization of photon is defined by
\begin{equation}\label{Eq:12}
P_\gamma=\frac{\sigma_+-\sigma_-}{\sigma_++\sigma_-}\,,
\end{equation}
where $\sigma_+$ and $\sigma_-$ are the capture cross section for
photons with positive and negative helicity, respectively.

From Eq.(\ref{Eq:12}), we can write the PV polarization of photon in
$nd\rightarrow$ $^3H\gamma$ as
\begin{eqnarray}\label{Eq:13}
P^{nd}_\gamma=
2\frac{Re\Big[{\mathcal{W}^{PC}}^\dagger(^2S_{\frac{1}{2}})\mathcal{W}^{PV}(^2S_{\frac{1}{2}})+\:{\mathcal{W}^{PC}}^\dagger(^4S_{\frac{3}{2}})\mathcal{W}^{PV}(^4S_{\frac{3}{2}})\Big]}{\big|\mathcal{W}^{PC}(^2S_{\frac{1}{2}})\big|^2+\big|\mathcal{W}^{PC}(^4S_{\frac{3}{2}})\big|^2},
\end{eqnarray}
where $\mathcal{W}^{PC}(X)$ and $\mathcal{W}^{PV}(X)$ are the PC
(M1) and PV (E1) amplitudes of $nd\rightarrow$ $^3H\gamma$ process
which are introduced and evaluated in Sects.\ref{PC nd radiative
capture} and \ref{PV nd radiative capture}, respectively.

We insert the results of the $\mathcal{W}^{PC}(X)$ and
$\mathcal{W}^{PV}(X)$ for both incoming doublet and quartet channels
in Eq.(\ref{Eq:13}). The calculated result of $P^{nd}_\gamma$ in
terms of PV LECs is given by
\begin{eqnarray}\label{Eq:032}
P^{nd}_\gamma=\big[\;0.26\,g^{(^3S_1-^1P_1)}\qquad\qquad\nonumber
\\-0.23\,g^{(^1S_0-^3P_0)}_{(\Delta
I=0)}\qquad\quad\;\;\nonumber \\+0.15\,g^{(^1S_0-^3P_0)}_{(\Delta
I=1)}\qquad\quad\;\;\nonumber
\\-1.08\,g^{(^3S_1-^3P_1)}\big]\times10^{3}.\,
\end{eqnarray}

\subsubsection{Photon asymmetries in $nd\rightarrow$ $^3H\gamma$}\label{Ag}
In the three-body system, other PV observables are the asymmetries
of the $\gamma$-emission. The equation
\begin{eqnarray}\label{Eq:14}
\frac{1}{\Gamma}\frac{d\Gamma}{d\,cos\theta}=1+O_\gamma\,cos\theta
\end{eqnarray}
represents the relation that gives us the asymmetry of photon. In
Eq.(\ref{Eq:14}) $O_\gamma$ represents e.g. $a^{nd}_\gamma$
($A^{nd}_\gamma$) which is the asymmetry of the outgoing photon with
respect to the neutron (deuteron) polarization in $nd\rightarrow$
$^3H\gamma$. $\Gamma$ is the process width and $\theta$ denotes the
angle between neutron (deuteron) polarization and outgoing photon
direction. By using Eq.(\ref{Eq:14}), we can see that
$a^{nd}_\gamma$ and $A^{nd}_\gamma$ are given in terms of the PC and
PV amplitudes of $nd\rightarrow$ $^3H\gamma$ by,
\begin{eqnarray}\label{Eq:15}
a^{nd}_\gamma=\frac{2}{3}Re\bigg[\sqrt{2}{\mathcal{W}^{PC}}^\dagger(^2S_{\frac{1}{2}})\mathcal{W}^{PV}(^4S_{\frac{3}{2}})+\sqrt{2}{\mathcal{W}^{PC}}^\dagger(^4S_{\frac{3}{2}})\mathcal{W}^{PV}(^2S_{\frac{1}{2}})\qquad\!\!
\nonumber
\\+\frac{5}{2}\:{\mathcal{W}^{PC}}^\dagger(^4S_{\frac{3}{2}})\mathcal{W}^{PV}(^4S_{\frac{3}{2}})-\:{\mathcal{W}^{PC}}^\dagger(^2S_{\frac{1}{2}})\mathcal{W}^{PV}(^2S_{\frac{1}{2}})\bigg]\qquad\nonumber \\ \bigg/\bigg[\big|\mathcal{W}^{PC}(^2S_{\frac{1}{2}})\big|^2
+\big|\mathcal{W}^{PC}(^4S_{\frac{3}{2}})\big|^2\bigg],\qquad\qquad\qquad\qquad\qquad
\end{eqnarray}
and
\begin{eqnarray}\label{Eq:015}
A^{nd}_\gamma=-Re\bigg[\sqrt{2}{\mathcal{W}^{PC}}^\dagger(^2S_{\frac{1}{2}})\mathcal{W}^{PV}(^4S_{\frac{3}{2}})+\sqrt{2}{\mathcal{W}^{PC}}^\dagger(^4S_{\frac{3}{2}})\mathcal{W}^{PV}(^2S_{\frac{1}{2}})\qquad\!\!\nonumber
\\-5\:{\mathcal{W}^{PC}}^\dagger(^4S_{\frac{3}{2}})\mathcal{W}^{PV}(^4S_{\frac{3}{2}})
-4\:{\mathcal{W}^{PC}}^\dagger(^2S_{\frac{1}{2}})\mathcal{W}^{PV}(^2S_{\frac{1}{2}})\bigg]\qquad\!\!\nonumber
\\ \bigg/\bigg[\big|\mathcal{W}^{PC}(^2S_{\frac{1}{2}})\big|^2
+\big|\mathcal{W}^{PC}(^4S_{\frac{3}{2}})\big|^2\bigg].\qquad\qquad\qquad\qquad\qquad
\end{eqnarray}

From Eqs.(\ref{Eq:15}) and (\ref{Eq:015}) and the results of PC and
PV amplitudes of $nd\rightarrow$ $^3H\gamma$, we obtain
\begin{eqnarray}\label{Eq:16}
a^{nd}_\gamma=\big[\;-0.51\,g^{(^3S_1-^1P_1)}\qquad\;\;\;\nonumber
\\+0.83\,g^{(^1S_0-^3P_0)}_{(\Delta
I=0)}\qquad\;\;\;\;\nonumber \\-0.47\,g^{(^1S_0-^3P_0)}_{(\Delta
I=1)}\qquad\;\;\;\;\nonumber
\\+1.36\,g^{(^3S_1-^3P_1)}\big]\times10^{3},\!
\end{eqnarray}
\begin{eqnarray}\label{Eq:00016}
A^{nd}_\gamma=\big[\;1.36\,g^{(^3S_1-^1P_1)}\qquad\quad\;\;\nonumber
\\-1.50\,g^{(^1S_0-^3P_0)}_{(\Delta
I=0)}\qquad\;\;\;\;\nonumber \\+0.94\,g^{(^1S_0-^3P_0)}_{(\Delta
I=1)}\qquad\;\;\;\;\nonumber
\\-4.47\,g^{(^3S_1-^3P_1)}\big]\times10^{3}.\!
\end{eqnarray}
We note that the contributions of the diagrams in Figs.\ref{Fig:PC
nd capture} and \ref{Fig:PV Nd capture} are calculated with the same
cutoff value, $\Lambda=700\,\textrm{MeV}$. The evaluation of the
photon asymmetries in $nd\rightarrow$ $^3H\gamma$ will be carried
out in Sect.\ref{Ag calculation}, however primarily, we need to
obtain the value of the PV LECs in the next section.

\section{The determination of the parity-violating low-energy coupling constants}\label{PV LECs}

The EFT($\pi\!\!\!\!/$) predicted relations for PV observables with
the capability of power counting provide a reliable calculation with
controlled theoretical errors. Therefore, the PV observables can be
introduced in terms of five PV LECs according to the formalism based
on EFT($\pi\!\!\!\!/$) framework. However, the lack of experimental
values for the PV observables are the main obstacle so for to
determine the PV LECs.

At the present circumstances, we intend to match the
EFT($\pi\!\!\!\!/$) calculated relations for the PV observables with
the DDH estimates in order to obtain the PV LECs.

\begin{table*}[tb]\centering \caption{The theoretical values of
five PV observables which are used in the determination of the PV
LECs.}
\label{tab:exp. PV data}       
\begin{tabular}{ccc}
\hline\noalign{\smallskip}

$\quad$PV observable$\quad$ & $\quad$Model used for matching$\quad$& $\quad$Value$\quad$  \\
\hline\hline\noalign{\smallskip}

$P^{np}_\gamma$ \cite{Song-L-G} &$\textrm{AV18+DDH-II}$&
$1.76\times10^{-8}$\\
 $A^{np}_\gamma$\cite{Song-L-G} & $\textrm{AV18+DDH-II}$&
$5.29\times10^{-8}$ \\
$\frac{d\phi^{np}}{dl}$ \cite{G-S-S}& DDH-best & $3.2\times 10^{-17}$ $\textrm{rad.fm}^{-1}$ \\
$\frac{d\phi^{nd}}{dl}$  \cite{Schiavilla-V-G-K-M}& \textrm{AV18+UIX/DDH-best} & $9.32\times 10^{-17}$ $\textrm{rad.fm}^{-1}$\\
$P^{nd}_\gamma$\cite{Song-L-G} &$\textrm{AV18+UIX/DDH-II}$&
$-7.30\times10^{-7}$ \\


\noalign{\smallskip}\hline\\

\end{tabular}
\end{table*}

The EFT($/\!\!\!\pi$) relations for $A^{np}_\gamma$ and
$P^{np}_\gamma$ observables at LO are \cite{schindler-springer}
\begin{eqnarray}\label{Eq:016}
A^{np}_\gamma=2m_N^2 \sqrt{\frac{\rho_d}{\pi}}\frac{1-\frac{\gamma_t
a_t}{3}}{k_1(1-\gamma_t a_s)}g^{(^3S_1-^3P_1)},
\end{eqnarray}
and
\begin{eqnarray}\label{Eq:0016}
P^{np}_\gamma=-2\sqrt{\frac{\rho_d}{\pi}}\frac{m_N^2}{k_1(1-\gamma_t
a_s)}\bigg[\Big(1-\frac{2}{3}\gamma_t a_s\Big)g^{(^3S_1-^1P_1)}
+\frac{\gamma_t
a_s}{3}\sqrt{\frac{r_0}{\rho_d}}\Big(g^{(^1S_0-^3P_0)}_{(\Delta
I=0)}-2\,g^{(^1S_0-^3P_0)}_{(\Delta I=2)}\Big)\bigg],\qquad\!\!\!
\end{eqnarray}
where $a_t$ is the $NN$ scattering length in $^3S_1$ channel. $r_0$
and $\rho_d$ denote the effective ranges in singlet and triplet $NN$
channels, respectively. We use the nucleon mass $m_N=938.918$ MeV,
the isovector nucleon magnetic moment $k_1=2.35294$, deuteron
binding momentum $\gamma_t=45.7025$ MeV, effective range of $NN$
singlet (triplet) channel $r_{0}=2.73$ fm ($\rho_{d}=1.764$ fm) and
scattering length in the singlet (triplet) channel $a_s=-23.714$ fm
($a_t=\frac{1}{\gamma_t}$), so we have
\begin{eqnarray}\label{Eq:17}
A^{np}_\gamma=\Big(4.102\,g^{(^3S_1-^3P_1)}\Big)\times10^{3},
\end{eqnarray}
\begin{eqnarray}\label{Eq:18}
P^{np}_\gamma=\!\Big(\!-28.699\,g^{(^3S_1-^1P_1)}+14.024\,\big[g^{(^1S_0-^3P_0)}_{(\Delta
I=0)}-2\,g^{(^1S_0-^3P_0)}_{(\Delta I=2)}\big]\Big)\times10^{3},
\end{eqnarray}

The EFT($/\!\!\!\pi$) results of $np$ and $nd$ spin rotations in
terms of the PV coupling constants at NLO are \cite{G-S-S},
\begin{eqnarray}\label{Eq:018}
\frac{1}{\rho}\frac{d\phi^{np}}{dl}=\Big(4.5\,\big[2g^{(^3S_1-^3P_1)}+g^{(^3S_1-^1P_1)}\big]
-18.5\, \big[g^{(^1S_0-^3P_0)}_{(\Delta
I=0)}-2\,g^{(^1S_0-^3P_0)}_{(\Delta
I=2)}\big]\Big)\,\textrm{rad.MeV}^{-2},
\end{eqnarray}
and
\begin{eqnarray}\label{Eq:0018}
\frac{1}{\rho}\frac{d\phi^{nd}}{dl}=\Big(8.0\,g^{(^3S_1-^3P_1)}+17.0\,g^{(^3S_1-^1P_1)}+2.3\big[3g^{(^1S_0-^3P_0)}_{(\Delta
I=0)}-2\,g^{(^1S_0-^3P_0)}_{(\Delta
I=1)}\big]\Big)\,\textrm{rad.MeV}^{-2},
\end{eqnarray}
where $\rho=0.04$ fm$^{-3}$ is the target density.

From Eqs.(\ref{Eq:032}) and (\ref{Eq:17}-\ref{Eq:0018}), we have
five relations which are sufficient for the determination of five PV
unknown coupling constants. For obtaining the PV coupling constants,
we use the results of PV observables based on the DDH method
presented in Table \ref{tab:exp. PV data}.

Our obtained results for $g^{(\bar{X}-\bar{Y})}$ are shown in Table
\ref{tab:result of g(X-Y)}. Our outcomes in Table \ref{tab:result of
g(X-Y)} are obtained using the results in Table \ref{tab:exp. PV
data} for $P^{np}_\gamma$, $A^{np}_\gamma$, $\frac{d\phi^{np}}{dl}$,
$\frac{d\phi^{nd}}{dl}$ and $P^{nd}_\gamma$ observables. In the next
section, we evaluate the $a^{nd}_\gamma$ and $A^{nd}_\gamma$ using
the values of $g^{(\bar{X}-\bar{Y})}$.

\begin{table*}[tb]\centering
\caption{Results of the obtained PV LECs in
$\textrm{MeV}^{-\frac{3}{2}}$ unit. Our outcomes are obtained using
the results of $P^{np}_\gamma$, $A^{np}_\gamma$,
$\frac{d\phi^{np}}{dl}$, $\frac{d\phi^{nd}}{dl}$ and $P^{nd}_\gamma$
observables which are introduced in Table \ref{tab:exp. PV data}.}
\label{tab:result of g(X-Y)}       
\begin{tabular}{ccccc}
\hline\hline\noalign{\smallskip}
$g^{(^3S_1-^1P_1)}$ & $g^{(^1S_0-^3P_0)}_{(\Delta I=0)}$ &
$g^{(^1S_0-^3P_0)}_{(\Delta I=1)}$ & $g^{(^1S_0-^3P_0)}_{(\Delta
I=2)}$ & $g^{(^3S_1-^3P_1)}$   \\
\noalign{\smallskip} \hline\noalign{\smallskip}

 \qquad$2.78\times10^{-12}$& \quad$4.71\times10^{-9}$&\quad
$2.41\times10^{-9}$&\quad
$2.35\times10^{-9}$ &\quad $1.29\times10^{-11}$ \\


 \noalign{\smallskip}\hline\hline
\end{tabular}
\end{table*}
\begin{table*}[tb]\centering
\caption{Comparison between different theoretical results for photon
asymmetries in $nd\rightarrow$ $^3H\gamma$ process. The row 9 shows
our EFT($\pi\!\!\!/$) results at LO. Our results in row 9 are
obtained using the determined values of the PV LECs which are
presented in Table \ref{tab:result of g(X-Y)}.}
\label{tab: Ag results}       
\begin{tabular}{ccc}
\hline\hline\noalign{\smallskip} Method & $a^{nd}_\gamma$ &
$A^{nd}_\gamma$ \\
\noalign{\smallskip} \hline\noalign{\smallskip}

DDH(RSC potential) \cite{desplanques-benayoun} & $0.61\times10^{-6}$  & $-1.40\times10^{-6}$   \\

DDH(SSC potential) \cite{desplanques-benayoun} & $0.81\times10^{-6}$  & $-1.60\times10^{-6}$ \\

DDH best values($\textrm{AV18+UIX/DDH-I}$) \cite{Song-L-G} & $3.30\times10^{-7}$  & $-8.23\times10^{-7}$\\
DDH best values($\textrm{AV18+UIX/DDH-II}$) \cite{Song-L-G} & $4.11\times10^{-7}$  & $-9.04\times10^{-7}$\\
DDH best values($\textrm{NijmII/DDH-II}$) \cite{Song-L-G} & $4.71\times10^{-7}$  & $-1.05\times10^{-6}$\\

4-parameter fits($\textrm{AV18+UIX/DDH-I}$) \cite{Song-L-G} & $1.97\times10^{-7}$  & $-1.81\times10^{-7}$ \\
4-parameter fits($\textrm{AV18+UIX/DDH-II}$) \cite{Song-L-G} & $4.14\times10^{-7}$  & $-4.09\times10^{-7}$ \\
4-parameter fits($\textrm{NijmII/DDH-II}$) \cite{Song-L-G} & $4.76\times10^{-7}$  & $-4.41\times10^{-7}$ \\

our EFT($\pi\!\!\!/$) & $2.82\times10^{-6}$  & $-4.87\times10^{-6}$ \\

Experiment  \cite{23 of shin-ando-hyun}& $\qquad(4.2\pm3.8)\times10^{-6}$ & - \\


 \noalign{\smallskip}\hline\hline
\end{tabular}
\end{table*}

\section{The result of photon asymmetries in $nd\rightarrow$
$^3H\gamma$ process}\label{Ag calculation} From Table
\ref{tab:result of g(X-Y)} and Eqs.(\ref{Eq:16}) and
(\ref{Eq:00016}), we can extract the values of the $a^{nd}_\gamma$
and $A^{nd}_\gamma$. We compare our EFT results of $a^{nd}_\gamma$
and $A^{nd}_\gamma$ with the previous theoretical calculations based
on the DDH model and the available experimental data in Table
\ref{tab: Ag results}. Our EFT($\pi\!\!\!/$) results in row 9 of
Table \ref{tab: Ag results} are evaluated using the outcomes shown
in Table \ref{tab:result of g(X-Y)}.

The sign and the order of $a^{nd}_\gamma$ results are rightly
comparable with the only available experimental data. There is no
experimental data for $A^{nd}_\gamma$, however, the results are in
the expected limits.

\section{Conclusion and Outlook} \label{conclusion}
In this paper, we have incorporated five relations for PV
observables in terms of PV LECs in the EFT($\pi\!\!\!/$) framework.
We obtain the PV LECs by matching these five relations onto the
corresponding results for PV observables which are calculated based
on the different DDH PV potential combinations.

The extraction of the results for $a^{nd}_\gamma$ and
$A^{nd}_\gamma$ have been done using the relationship between
corresponding observables and the PV low-energy coupling constants.

The asymmetries from $\overrightarrow{\gamma}$$^3H\rightarrow nd$ or
$\overrightarrow{\gamma}$$^3He\rightarrow pd$ and the longitudinal
asymmetry in $\overrightarrow{p}d$ scattering are the other
theoretical analyses which could be studied in the
EFT($\pi\!\!\!\!/$) framework using the determined PV LECs.

The model independent EFT($\pi\!\!\!/$) predictions for observables
and particularly PV observables in terms of PV LECs require
principally to be fixed by experimental data. However, this task
does not look feasible in the nearest future.

\begin{acknowledgements}
This work was supported by the research council of the University of
Tehran.
\end{acknowledgements}

\setcounter{section}{0} \setcounter{subsection}{0}
\setcounter{equation}{0}
\renewcommand{\theequation}{\Alph{section}.\arabic{equation}}
\renewcommand{\thesection}{Appendix}
\renewcommand{\thesubsection}{Appendix \Alph{section}.\arabic{subsection}}
\section{Introduction of the diagrams used in the text} \label{Appendix A}

\begin{figure*}[tb]\centering
\includegraphics*[width=13.5cm]{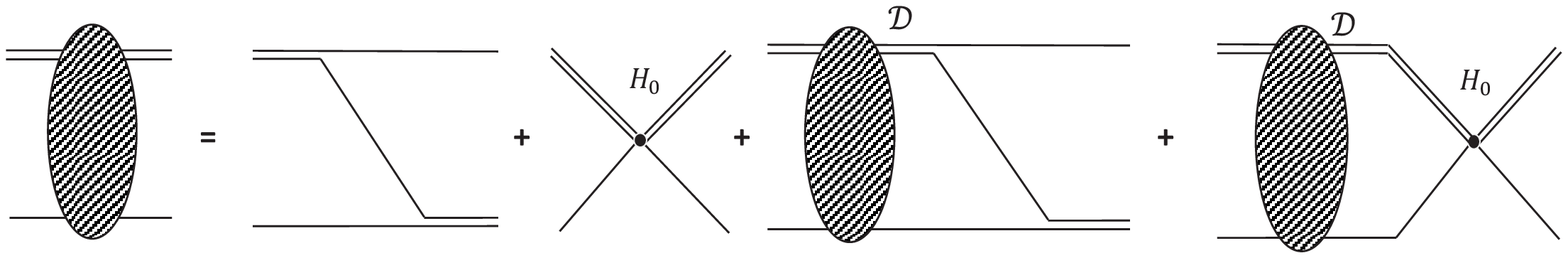}
\caption{\label{Fig:PC nd scattering}The Faddeev equation for the
$Nd$ scattering at the leading order. All notations are the same as
the previous figures.}
\end{figure*}

The PC $Nd$ scattering amplitude is used for calculating the PC and
PV amplitudes of $nd\rightarrow$ $^3H\gamma$ process. In
Figs.\ref{Fig:PC nd capture} and \ref{Fig:PV Nd capture}, the PC
amplitude of $Nd$ scattering at LO is introduced by the dashed oval.
The Faddeev equation that calculates the PC $Nd$ scattering
amplitude is schematically shown in Fig.\ref{Fig:PC nd scattering}.
The detail calculation process for solving this Faddeev equation in
an arbitrary channel in the cluster-configuration space is
previously reported in \cite{20 of sadeghi-bayegan}. We emphasize
that the normalized triton wave function, which is introduced in the
text with the dashed half-oval, is obtained by solving the
homogeneous part of Faddeev equation of the PC $Nd$ scattering
\cite{moeini-bayegan}.

\begin{figure*}[tb]\centering
\includegraphics*[width=12cm]{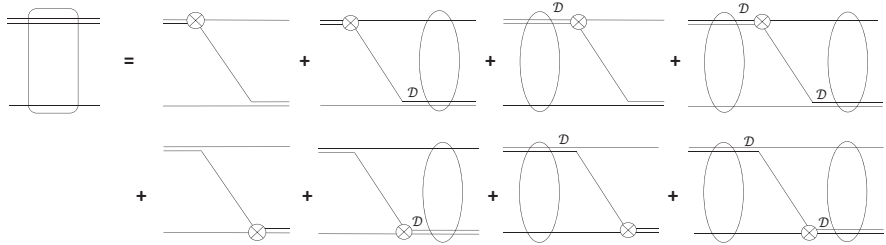} \caption{\label{Fig:PV Nd scattering}The PV $Nd$ scattering diagrams at LO. The
dashed rectangular denotes the PV $Nd$ scattering amplitude. Circle
with a cross indicates the PV $dNN$ vertex. Remaining notations are
the same as the previous figures.}
\end{figure*}

\begin{figure*}[tb]\centering
\includegraphics*[width=15.5cm]{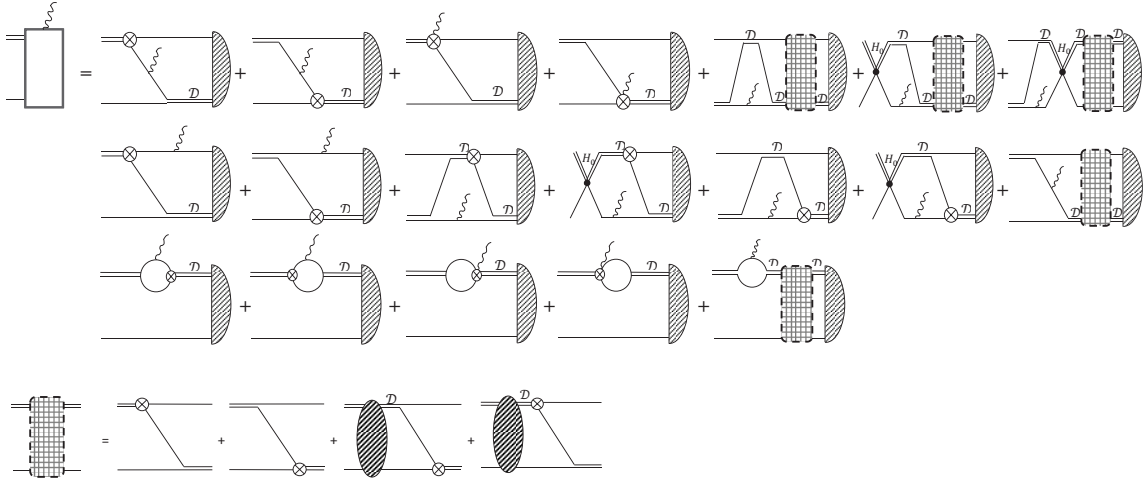} \caption{\label{Fig:additional PV Nd capture}
The box with wavy line which is used in Fig.\ref{Fig:PV Nd capture}.
Circle with a cross and wavy line is the PV E1
photon-dibaryon-nucleon-nucleon ($\gamma dNN$) vertex. The dashed
rectangular with dashed line around it is the contribution of the
some of the PV $Nd$ scattering diagrams which are introduced in the
last line. All notations are the same as the Fig.\ref{Fig:PV Nd
capture}.}
\end{figure*}

In the text, Fig.\ref{Fig:PV Nd capture} shows the diagrams which
participate in the PV amplitude of $nd\rightarrow$ $^3H\gamma$
process. The diagrams making the dashed rectangular and the box with
the wavy line of Fig.\ref{Fig:PV Nd capture} are introduced in
Figs.\ref{Fig:PV Nd scattering} and \ref{Fig:additional PV Nd
capture}, respectively. In Fig.\ref{Fig:additional PV Nd capture},
the dashed rectangular with the dashed line around it is the sum of
the PV $Nd$ scattering diagrams which have no PC $Nd$ scattering in
the right hand side. The last line of Fig.\ref{Fig:additional PV Nd
capture} shows these diagrams.




\end{document}